\documentclass[twocolumn,showpacs,preprintnumbers,amsmath,amssymb]{revtex4}

\usepackage{graphicx}
\usepackage{dcolumn}
\usepackage{bm}

\begin{document}


\title{Algebraic approach to solve $t\bar{t}$ dilepton equations}

\author{Lars Sonnenschein}

\affiliation{RWTH Aachen University, III. Physikalisches Institut A, 52056 Aachen}

\date{\today}

\begin{abstract}
  The set of non-linear equations describing the Standard Model kinematics of the top quark antiqark production 
  system in the dilepton decay channel has at most a four-fold 
  ambiguity due to two not fully reconstructed neutrinos.
  Its most precise solution is of major importance for measurements of top quark properties like
  the top quark mass and $t\bar{t}$ spin correlations.
  Simple algebraic operations allow to transform the non-linear equations into a system of two 
  polynomial equations with two unknowns. These two polynomials of multidegree eight can in turn be analytically 
  reduced to one polynomial with one unknown by means of resultants. 
  The obtained univariate polynomial is of degree sixteen.
  The number of its real solutions 
  is determined analytically by means of Sturm's theorem, 
  which is as well used to
  isolate each real solution into a unique pairwise disjoint interval.
  The solutions are polished by seeking the sign 
  change of the polynomial in a given interval through binary bracketing.


\end{abstract}

\maketitle

\section{Introduction}

\noindent
In 1992 R. H. Dalitz and G. R. Goldstein have published a numerical method based on
geometrical considerations to solve the system of equations describing the kinematics
of the $t\bar{t}$ decay in the dilepton channel \cite{Dalitz1992}\cite{Goldstein1992}. 
The problem of two not fully 
reconstructed neutrinos - only the transverse components of the vector sum of their missing energy 
can be measured - leads to a system of equations which consists of as many equations
as there are unknowns. Thus it is straight forward to solve the system of equations directly
in contrast to a kinematic fit which would be appropriate in the case of an over-constrained 
problem or integration over the phase space of degrees of freedom 
in case of an under-constrained problem. Each of the two neutrinos contributes a two-fold 
ambiguity to the solution of the system of equations which end up to an over-all
ambiguity of degree four. On top of those ambiguities which dilute the significance
of top quark property measurements in the dilepton channel, 
reconstructed objects do typically not coincide
with their corresponding particles which reduces the significance further.
Thus it is not only important to solve the system of equations but also to 
compare its solutions to the particle momenta of simulated events.

Next section the system of equations is introduced,
followed by a description of the algebraic solution and its implementation as algorithm. 
Subsequently the performance of the numerical implementation is discussed.  

\section{$t\bar{t}$ dilepton kinematics}

\noindent
The system of equations describing the kinematics of $t\bar{t}$
dilepton events can be expressed by the two linear and six non linear
equations
\begin{eqnarray} 
\label{initialequations} 
\nonumber 
E_x\!\!\!\!\!\!/ \;\;& = & p_{\nu_x} + p_{\bar{\nu}_x} \\ \nonumber
E_y\!\!\!\!\!\!/ \;\;& = & p_{\nu_y} + p_{\bar{\nu}_y} \\ \nonumber
E_{\nu}^2 & = & p_{\nu_x}^2+p_{\nu_y}^2+p_{\nu_z}^2+m_{\nu}^2 \\ \nonumber 
E_{\bar{\nu}}^2 & = & p_{\bar{\nu}_x}^2+p_{\bar{\nu}_y}^2+p_{\bar{\nu}_z}^2+m_{\bar{\nu}}^2 \\ \nonumber 
m_{W^+}^2 \!\! & = & (E_{\ell^+}+E_{\nu})^2-(p_{\ell^+_x}+p_{\nu_x})^2 \\ \nonumber
 & & -(p_{\ell^+_y}+p_{\nu_y})^2-(p_{\ell^+_z}+p_{\nu_z})^2 \\
m_{W^-}^2 \!\! & = & (E_{\ell^-}+E_{\bar{\nu}})^2-(p_{\ell^-_x}+p_{\bar{\nu}_x})^2 \\ \nonumber
 & & -(p_{\ell^-_y}+p_{\bar{\nu}_y})^2-(p_{\ell^-_z}+p_{\bar{\nu}_z})^2 \\ \nonumber
m_t^2 \; & = & (E_b+E_{\ell^+}+E_{\nu})^2-(p_{b_x}+p_{\ell^+_x}+p_{\nu_x})^2 \\ \nonumber
 & & -(p_{b_y}+p_{\ell^+_y}+p_{\nu_y})^2-(p_{b_z}+p_{\ell^+_z}+p_{\nu_z})^2 \\ \nonumber
m_{\bar{t}}^2 \; & = & (E_{\bar{b}}+E_{\ell^-}+E_{\bar{\nu}})^2-(p_{\bar{b}_x}+p_{\ell^-_x}+p_{\bar{\nu}_x})^2 \\ \nonumber
 & & -(p_{\bar{b}_y}+p_{\ell^-_y}+p_{\bar{\nu}_y})^2-(p_{\bar{b}_z}+p_{\ell^-_z}+p_{\bar{\nu}_z})^2 \; . 
\end{eqnarray}
The $z$-axis is here assumed to be parallel orientated to the beam axis while 
the $x$- and $y$-coordinates span the transverse plane.
The first two equations relate the projection of the missing transverse energy
onto one of the transverse axes ($x$ or $y$) to the sum of the neutrino and 
antineutrino momentum components belonging to the same projection.
The next two equations relate the energy of the neutrino and antineutrino
with their momenta. 
Finally four non linear equations describe the $W$ boson and top quark
(antiquark) mass constraints by relating the invariant masses to the
energy and momenta of their decay particles via relativistic 4-vector
arithmetics.

\section{Algebraic solution}

\noindent
This system of equations can be reduced to four equations
by simply substituting in the last four equations the neutrino and
antineutrino energies by the third and fourth equations
and substituting the antineutrino transverse momenta by the first two 
equations solved to these momenta. In this way the four unknowns $p_{\nu_x}$,
$p_{\nu_y}$, $p_{\nu_z}$ and $p_{\bar{\nu}_z}$ are left. One pair of equations,
describing the $t\rightarrow bW^+\rightarrow b\ell^+\nu_{\ell}$ parton branch 
of the event, depends on $p_{\nu_z}$ while the other pair of equations, 
describing the $\bar{t}\rightarrow \bar{b}W^-\rightarrow \bar{b}\ell^-\bar{\nu}_{\ell}$ parton branch
of the event, depends on $p_{\bar{\nu}_z}$. 
By means of ordinary algebraic operations both pairs can be solved to
the longitudinal neutrino and antineutrino momentum $p_{\nu_z}$ and $p_{\bar{\nu}_z}$
respectively. The equations can be written in the form
\begin{eqnarray} \label{pzequations} \nonumber
p_{\nu_z}=a_1\pm\sqrt{a_1^2+a_2} \\[-1ex]
\\[-1ex] \nonumber 
p_{\nu_z}=b_1\pm\sqrt{b_1^2+b_2} \nonumber
\end{eqnarray}  
for the top quark parton branch and
\begin{eqnarray} \label{pzbarequations} \nonumber
p_{\bar{\nu}_z}=c_1\pm\sqrt{c_1^2+c_2} \\[-1ex] 
 \\[-1ex] \nonumber
p_{\bar{\nu}_z}=d_1\pm\sqrt{d_1^2+d_2} 
\end{eqnarray}  
for the anti-top quark parton branch
with the coefficients
\begin{eqnarray} \label{pzcoefficients} \nonumber
a_1&=&a_{11}\!+a_{12}p_{\nu_x}\!+a_{13}p_{\nu_y} \\[-2ex] 
\\[0.7ex] \nonumber
a_2&=&a_{21}\!+a_{22}p_{\nu_x}\!+a_{23}p_{\nu_y}\!+a_{24}p_{\nu_x}^2\!+a_{25}p_{\nu_x}p_{\nu_y}\!+a_{26}p_{\nu_y}^2
\end{eqnarray}
and $b$ equivalent for the first pair of equations (\ref{pzequations}).
For the second pair of equations (\ref{pzbarequations}) holds analogically
\begin{eqnarray} \label{pzbarcoefficients} \nonumber
c_1&=&c_{11}+c_{12}p_{\bar{\nu}_x}\!+c_{13}p_{\bar{\nu}_y} \\[-2ex] 
\\[0.7ex] \nonumber
c_2&=&c_{21}+c_{22}p_{\bar{\nu}_x}\!+c_{23}p_{\bar{\nu}_y}\!+c_{24}p_{\bar{\nu}_x}^2\!+c_{25}p_{\bar{\nu}_x}p_{\bar{\nu}_y}\!+c_{26}p_{\bar{\nu}_y}^2
\end{eqnarray}
and $d$ equivalent. 
The explicit expressions in terms of the initial 
equations (\ref{initialequations}) are given in the appendix.
After equating both equations of each pair there remain
two equations with the two unknowns $p_{\nu_x}$ and $p_{\nu_y}$.

Again by means of ordinary algebraic operations the two non linear equations 
can be transformed into two polynomials of multi-degree eight.
To solve these two polynomials to $p_{\nu_x}$ the resultant with respect to the
neutrino momentum $p_{\nu_y}$ is computed as follows.
The coefficients and monomials of the two polynomials are rewritten in such a way
that they are ordered in powers of $p_{\nu_y}$ like
\begin{eqnarray} \nonumber \label{2poly}
f&=&f_1p_{\nu_y}^4+f_2p_{\nu_y}^3+f_3p_{\nu_y}^2+f_4p_{\nu_y}+f_5 \\ 
\\ \nonumber
g&=&g_1p_{\nu_y}^4+g_2p_{\nu_y}^3+g_3p_{\nu_y}^2+g_4p_{\nu_y}+g_5 \\ \nonumber 
\end{eqnarray} 
where $f$ and $g$ are polynomials of the remaining unknowns 
$p_{\nu_x}$, $p_{\nu_y}$ and the coefficients $f_m$, $g_n$ are 
univariate polynomials of $p_{\nu_x}$.
The resultant can then be obtained by computing the determinant of the 
Sylvester matrix
\begin{eqnarray} 
\mbox{Res}(p_{\nu_y}) & \!\! = \! & \! \mbox{Det} \! \left(
\begin{array}{cccccccc}
 f_1 & & & & g_1 & & & \\
 f_2 & f_1 & & & g_2 & g_1 & &  \\
 f_3 & f_2 & f_1 & & g_3 & g_2 & g_1 & \\
 f_4 & f_3 & f_2 & f_1 & g_4 & g_3 & g_2 & g_1 \\
 f_5 & f_4 & f_3 & f_2 & g_5 & g_4 & g_3 & g_2\\
 & f_5 & f_4 & f_3 & & g_5 & g_4 & g_3 \\
 & & f_5 & f_4 & & & g_5 & g_4 \\
 & & & f_5 & & & & g_5 \\
\end{array}
\right) = 0 \;\;\;\;\;\;\; \\ \nonumber
\end{eqnarray}
which is equated to zero. The omitted elements of the matrix are identical
to zero. Since each element in the matrix is a polynomial itself
the evaluation is very elaborative.
There are two ways to compute the determinant in practice.
The more elegant way from a programming technical point of view is to invoke 
recursively a function which computes 
subdeterminants and consists of a very limited number of lines. 
Unfortunately it turns out that this approach is too 
time consuming. The other way is to let Maple \cite{Maple2004} 
compute and optimize the determinant as a function of the unknown $p_{\nu_x}$ 
and implement it. This way the code grows orders of 
magnitude in size but on the other hand the evaluation speeds up by orders of magnitude.

The resultant is a univariate polynomial of the form
\begin{eqnarray} \label{poly16} \nonumber
0 & = & h_1p_{\nu_x}^{16}+h_2p_{\nu_x}^{15}+h_3p_{\nu_x}^{14}+h_4p_{\nu_x}^{13}+h_5p_{\nu_x}^{12}+h_6p_{\nu_x}^{11} \\
 & + & h_7p_{\nu_x}^{10}+h_8p_{\nu_x}^9+h_9p_{\nu_x}^8+h_{10}p_{\nu_x}^7+h_{11}p_{\nu_x}^6 \\ \nonumber
& + & h_{12}p_{\nu_x}^5+h_{13}p_{\nu_x}^4+h_{14}p_{\nu_x}^3+h_{15}p_{\nu_x}^2+h_{16}p_{\nu_x}+h_{17}  \\ \nonumber 
\end{eqnarray}  
with the remaining unknown $p_{\nu_x}$. It is of degree 16 and analytical solutions
of general univariate polynomials are only known until degree four.
Abel's impossibility theorem and Galois demonstrated that a univariate polynomial
of degree five can in general not be solved analytically 
with a finite number of additions, subtractions, multiplications, 
divisions, and root extractions \cite{Weisstein}.
Thus from here on the solutions of the univariate polynomial (\ref{poly16})
have to be obtained by different means.
In principle the problem can be reduced to an Eigenvalue problem.
Unfortunately, in practice it turns out that the implementation of the Eigenvalue
package in Root \cite{Root2004} gives only reasonable solutions for univariate
polynomials of degree 14 and below.
Finally the number of solutions is obtained analytically by applying 
Sturm's theorem \cite{Groebner1993} which consists of building a sequence
of univariate polynomials $h(p_{\nu_x}), h'(p_{\nu_x}), h_a(p_{\nu_x}), h_b(p_{\nu_x}),..., h_m(p_{\nu_x})=$ const.,
where $h'$ is the first derivative of the univariate polynomial $h$ with
respect to $p_{\nu_x}$ and the following polynomials are the remainders
of a long division of their immediate left neighbour polynomial
divided by the next left neighbour polynomial.
The sequence ends when the last polynomial is a constant.
In the case the constant vanishes, the initial polynomial has at least 
one multiple real root which can be splitted by long division through the
last non constant polynomial in the Sturm sequence.
In this case one solution is already known.
The sequence is evaluated at two neutrino momenta $p_{\nu_{x_{1,2}}}$
(initially at the kinematic limits) and the difference
between the number of sign changes of the evaluated sequence at the 
two interval limits is determined. 
The obtained quantity corresponds to the number 
of real solutions in the given interval.

This means that the theorem of Jacques Charles Fran\c{c}ois Sturm 
- which he has proven in 1829 \cite{Wikipedia} -
is extremely powerful since in the case of no real solutions no time  
needs to be spent for the unsuccessful attempt to find one.

To reduce numerical inaccuracies, all polynomial evaluations 
are applied using Horner's rule which factors out powers
of the polynomial variable $p_{\nu_x}$ \cite{Horner1998}.    
Further the solutions are separated by applying Sturm's theorem
with varying interval boundaries.
Once the solutions are separated in unique pairwise disjoint intervals
they are polished by binary bracketing exploiting the knowledge about
the sign change at the root in the given interval.
This is possible since it is guaranteed that
there is only one single solution in a given interval per construction
\begin{figure}[b]
\hspace*{1ex}\includegraphics[width=9cm]{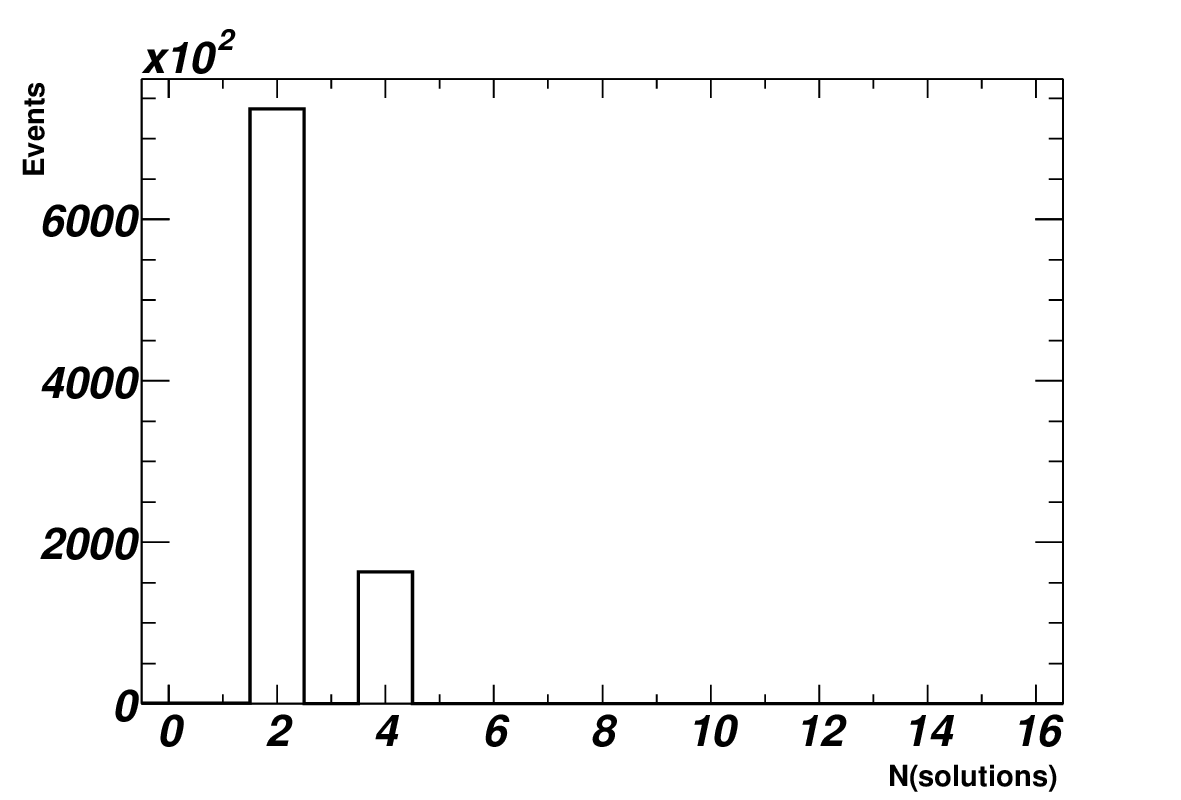}
\caption{\label{Nsol} Number of solutions per event.
}
\end{figure}
\begin{figure}[b]
\vspace*{-5ex}
\hspace*{-5ex}\includegraphics[width=9cm]{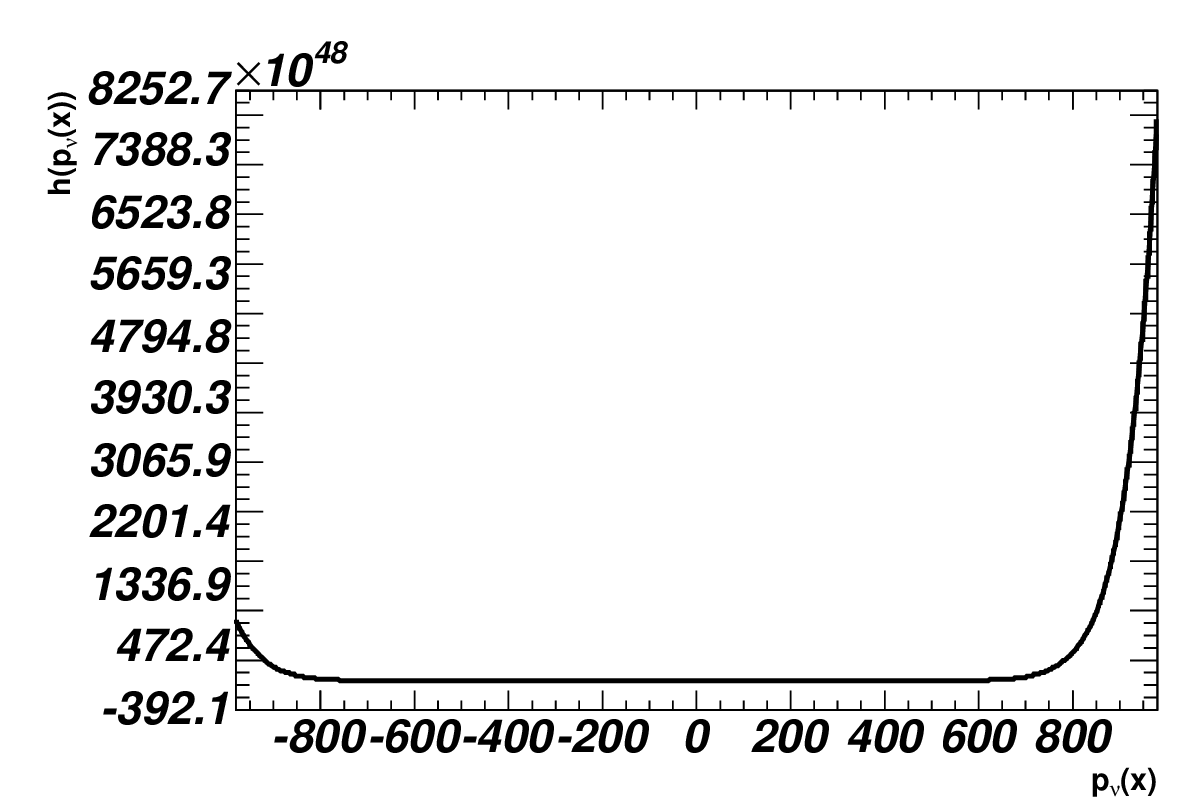}
\hspace*{-5ex}\includegraphics[width=9cm]{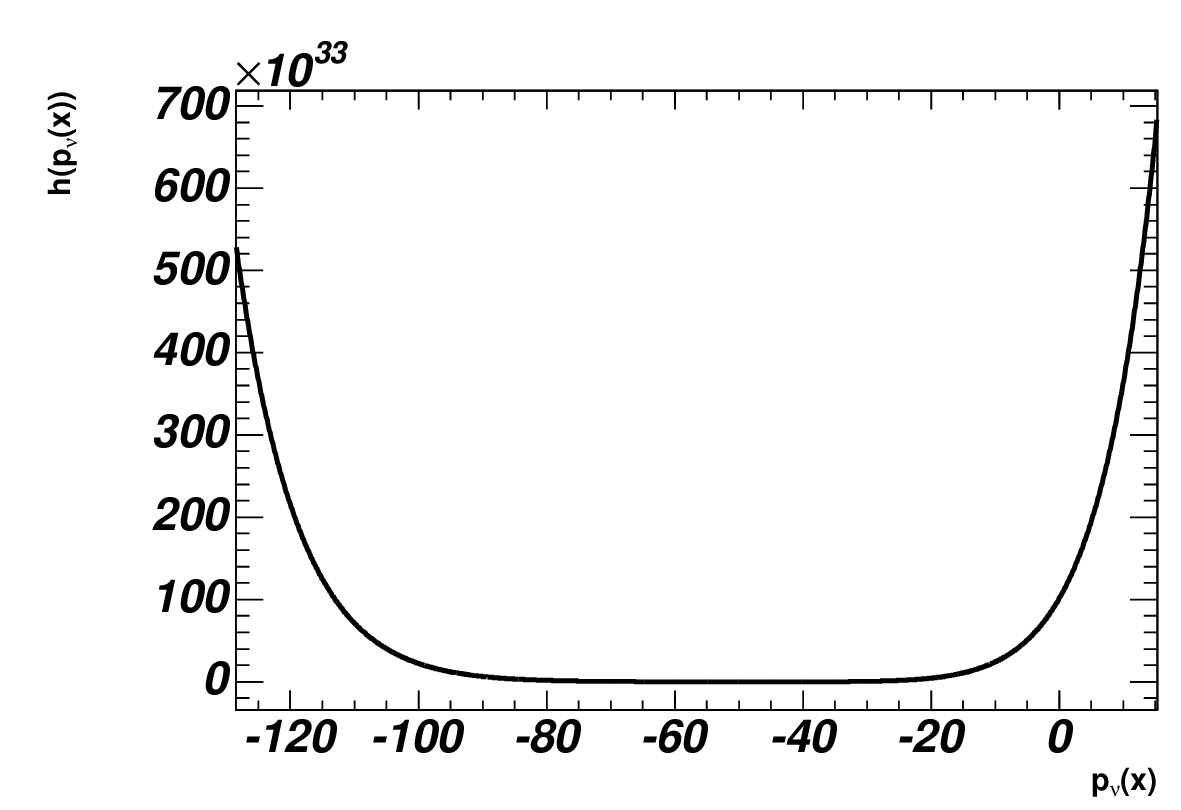}
\hspace*{-5ex}\includegraphics[width=9cm]{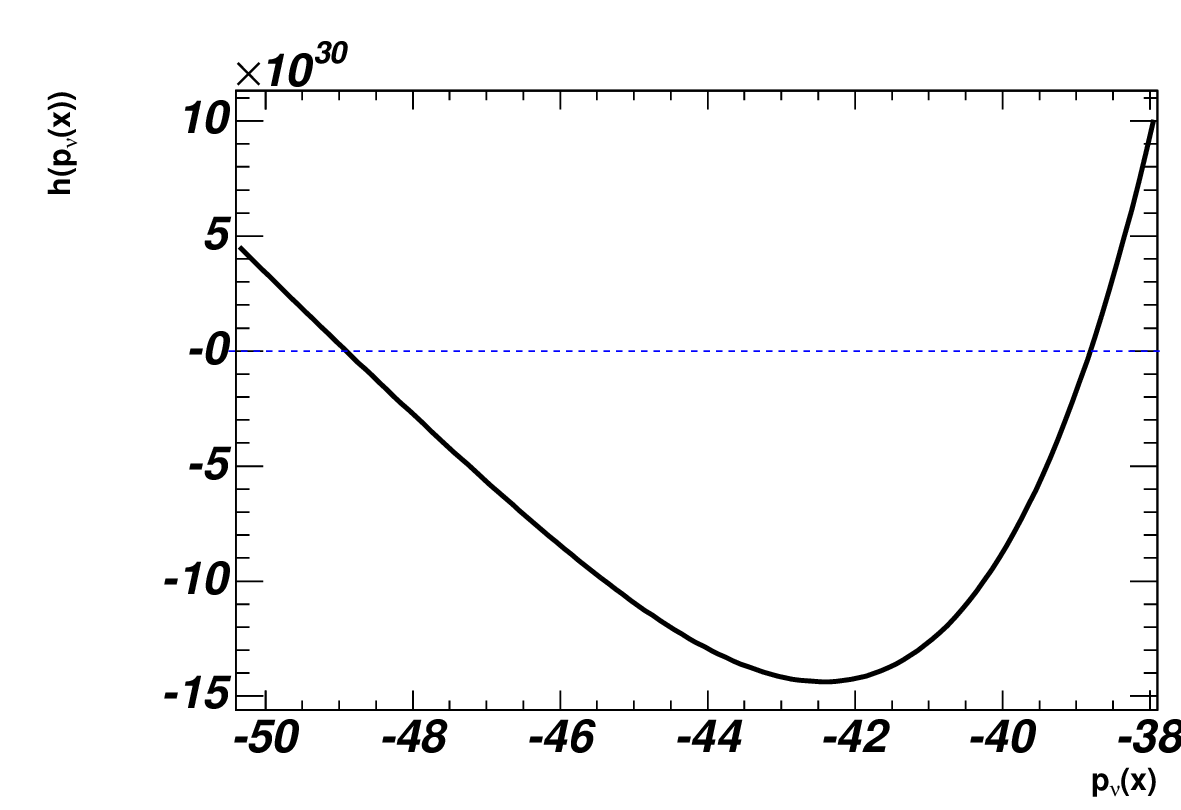}
\caption{\label{unif} A typical univariate polynomial of degree 16
whose real roots in $p_{\nu_x}$ are solutions of the initial system 
of equations describing the $t\bar{t}$ dilepton kinematics.
From top to bottom the plots are zoomed around the interesting
$p_{\nu_x}$ range of the abscissa where two solutions are located.
}
\vspace*{-4ex}
\end{figure}
(Now one could turn the way to solve a given Eigenvalue problem
the other way around and use the Sturm sequence
to solve the characteristic polynomial to obtain the Eigenvalues).

The neutrino and antineutrino masses are assumed to be zero in good approximation
in the following. They have been kept in the equations for the sake of completeness 
since the same set of equations can be exploited in search for new physics with the same decay topology
including invisible massive particles.

Once the solutions are found - most frequently 
there are two but never more than four (see fig. \ref{Nsol}) - they can be 
inserted in equations (\ref{2poly}).
Such that these equations reduce to two univariate polynomials of degree four 
which in turn can be solved analytically to $p_{\nu_y}$ 
with a four fold ambiguity.
The ambiguities can be eliminated in requiring the roots
of these two polynomials to coincide 
since both equations have to be satisfied simultaneously.
$p_{\bar{\nu}_x}$ and $p_{\bar{\nu}_y}$ can be simply determined with help 
of the first two equations in (\ref{initialequations}). 
To determine the longitudinal neutrino and antineutrino momenta $p_{\nu_z}$ and
$p_{\bar{\nu}_z}$ the equations (\ref{pzequations}) and (\ref{pzbarequations})
can be evaluated respectively. Again the two-fold ambiguity, here due to the 
square root sign, can be resolved in requiring the solutions to coincide
simultaneously for both equations of one parton branch.

\section{Performance of the method}

\noindent
The univariate polynomial of $p_{\nu_x}$
is in general very shallow around zero over a broad range of neutrino momenta
in comparison to its maximal values in the allowed kinematic range
as can be concluded from the first two graphs in fig. \ref{unif}. 
Here the kinematic range has been restricted to a centre of mass energy
of 1.96~TeV, assuming the Tevatron proton anti-proton collider 
which has been set up in the Monte-Carlo 
event generator PYTHIA 6.220 \cite{PYTHIA2001} used here.
Cross checks at a centre of mass energy of 14~TeV assuming the LHC
proton proton collider environment confirm that the performance is
independent of particular collider settings.
Only when in the graphs the area of the abscissa is zoomed very close to the 
solutions they can be recognised by eye. At this level the ordinate has 
already been magnified by 20 orders of magnitude.
This explains why it is in general so difficult to find any solutions 
with numerical methods. 

\begin{figure}[b]
\hspace*{-1ex}\includegraphics[width=9cm]{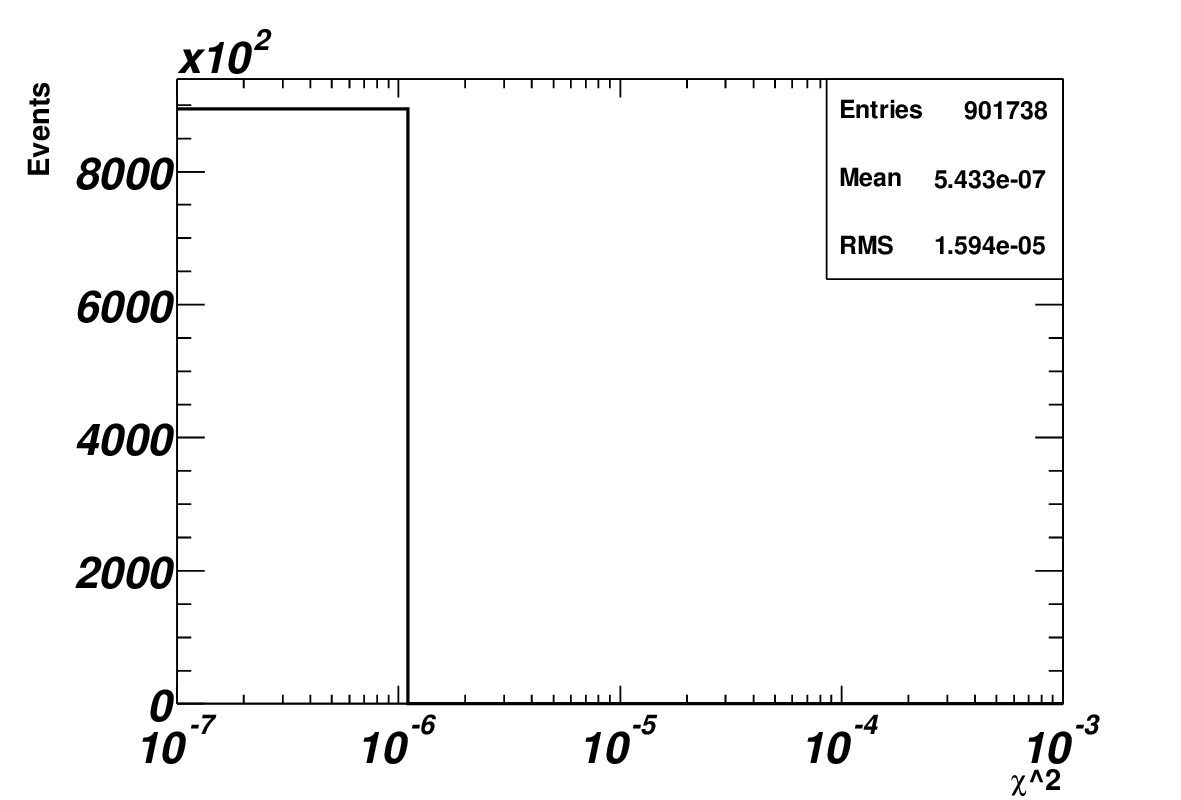}
\caption{\label{DataSolChi2} Solution $\chi^2$ defined as the difference
between solved and generated neutrino momenta, added in quadrature,
for the closest solution of each event.
}
\end{figure}

In 99.9\% of the events a solution can be found which 
is shown in the number of solutions per event distribution of fig. \ref{Nsol}. 
The neutrino momenta $p_{\nu}^{sol}$ of the solutions are compared to the generated ones
$p_{\nu}^{gen}$ by defining a metric $\chi$ through
\begin{eqnarray} \nonumber
\chi^2 & = & (p_{\nu_x}^{gen}\!-p_{\nu_x}^{sol})^2 +
	(p_{\nu_y}^{gen}\!-p_{\nu_y}^{sol})^2 +
	(p_{\nu_z}^{gen}\!-p_{\nu_z}^{sol})^2  \\
	\\ \nonumber
	& + & (p_{\bar{\nu}_x}^{gen}\!-p_{\bar{\nu}_x}^{sol})^2 +
	(p_{\bar{\nu}_y}^{gen}\!-p_{\bar{\nu}_y}^{sol})^2 +
	(p_{\bar{\nu}_z}^{gen}\!-p_{\bar{\nu}_z}^{sol})^2 \; .
\end{eqnarray}
%
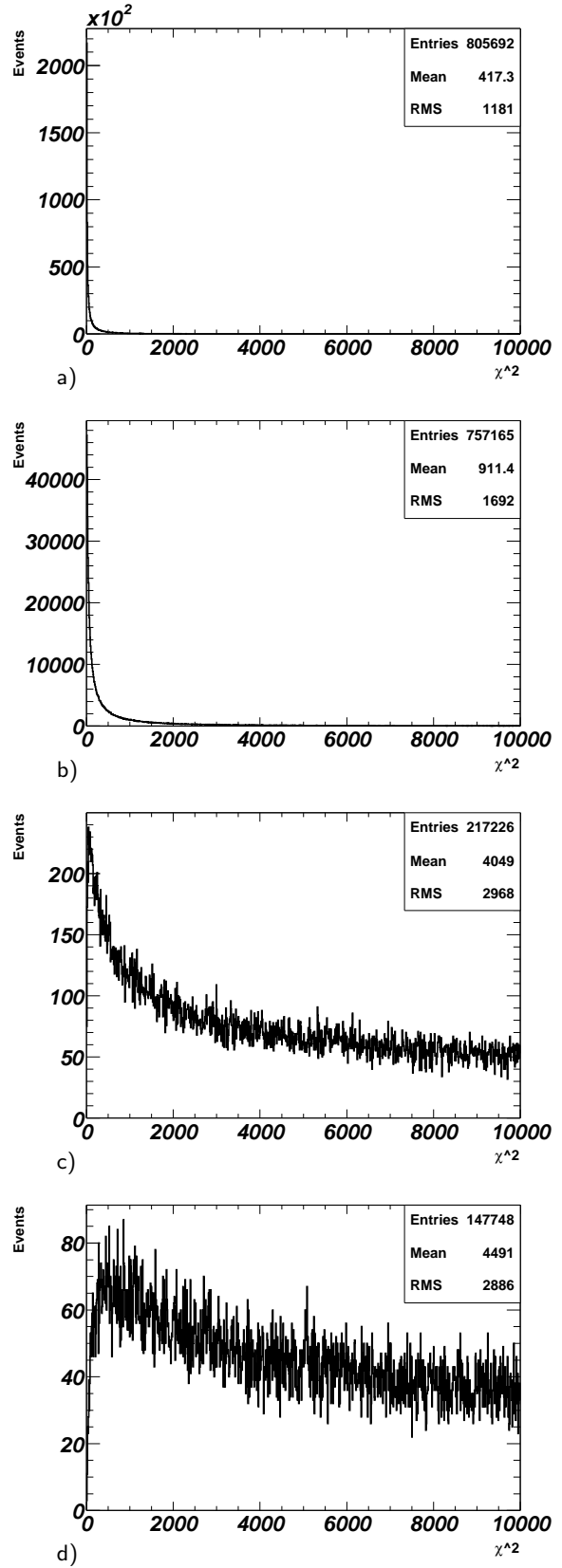
\begin{figure}[p]
\vspace*{8.5ex}
\setlength{\unitlength}{1mm}
\begin{picture}(50,200)
\put(-20,162){\includegraphics[width=8.1cm]{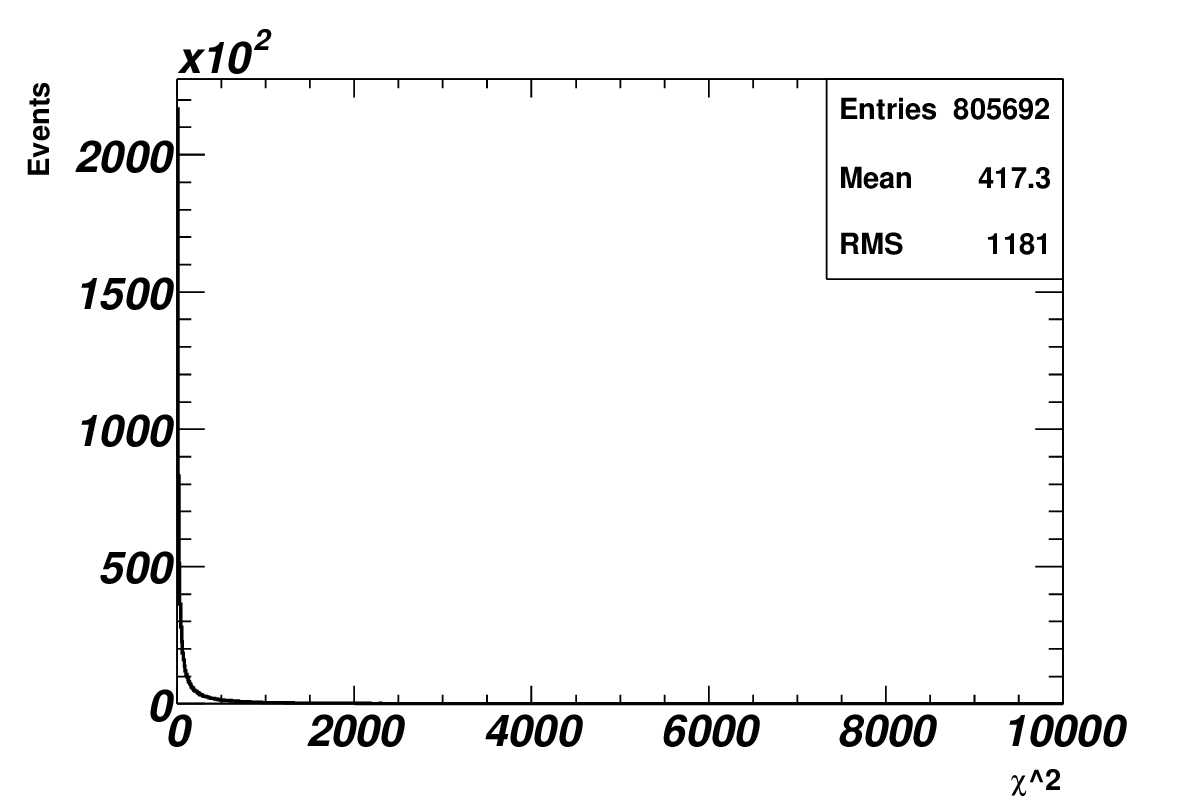} }
\put(-20,107){\includegraphics[width=8.1cm]{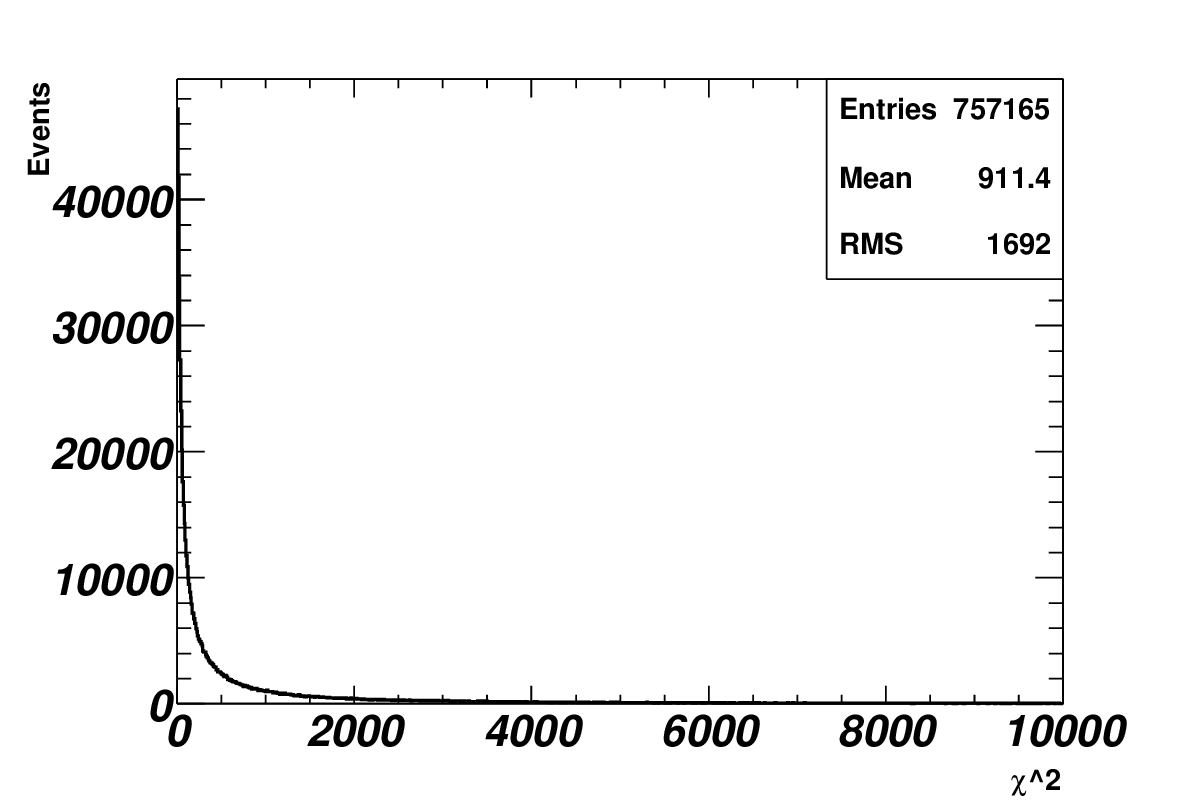} }
\put(-20,52){\includegraphics[width=8.1cm]{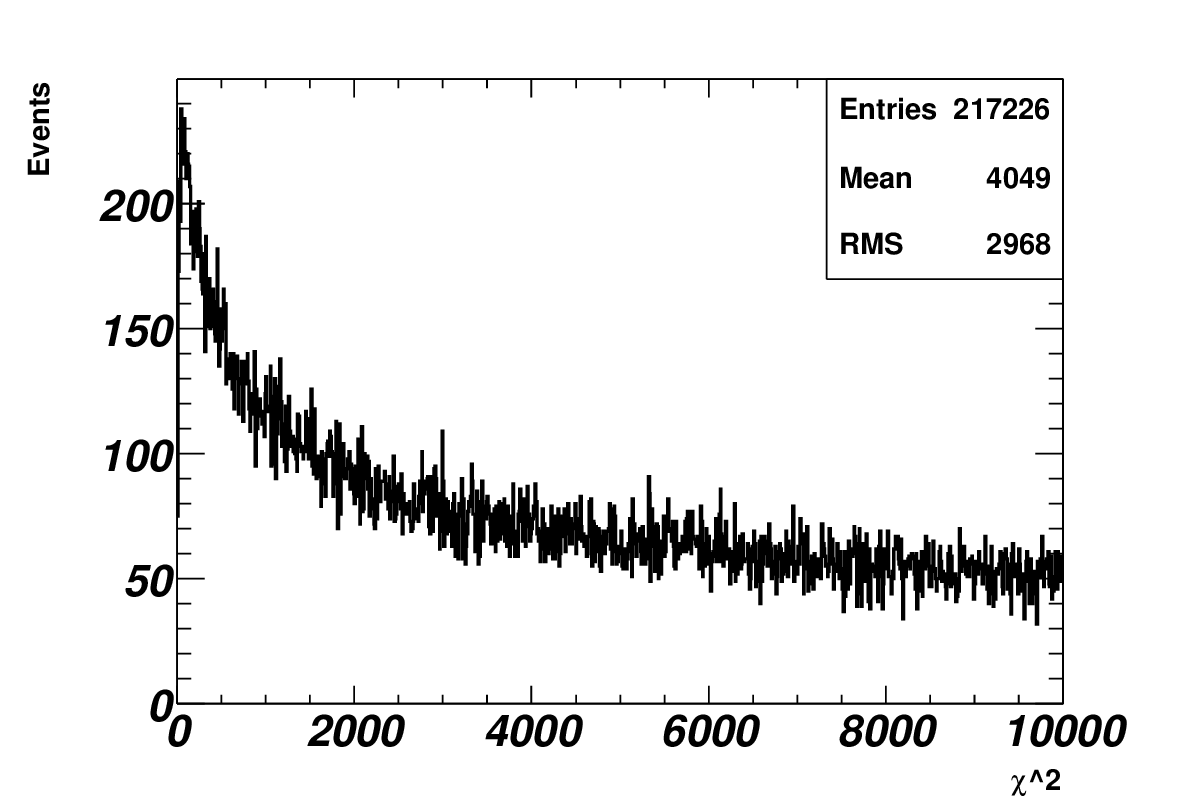} }
\put(-20,-3){\includegraphics[width=8.1cm]{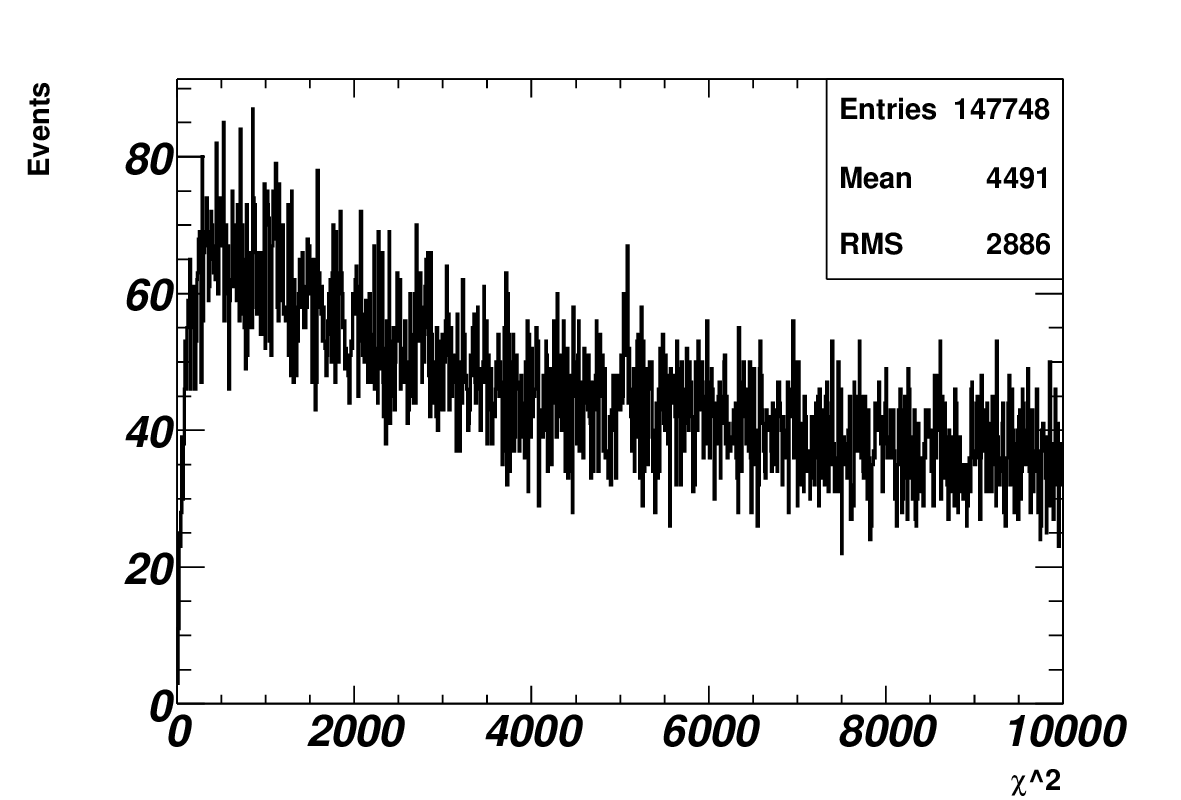} }
\put(-12,161.5){\sf a)}
\put(-12,106.5){\sf b)}
\put(-12,51.5){\sf c)}
\put(-12,-3.5){\sf d)}
\end{picture}
\vspace*{0.8ex}
\caption{\label{DataUnmatchEvtLeastSolChi2} Minimal solution $\chi^2$ per event. 
The first two plots differ in parton information entered into the solving procedure:
a) top quark pole and $W$ boson off-shell masses, b) top quark and $W$ boson pole masses.
The two lower plots show $\chi^2$ distributions of reconstructed events, 
considering both $b$ jet permutations with reconstructed jets in c) and 
additionally smearing applied to jets and leptons in d).
}
\end{figure}
\begin{table}[t]
\begin{tabular}{|l|c|c|c|} \cline{2-4}
 \multicolumn{1}{c|}{} & \multicolumn{3}{c|}{Solution} \\ \hline
Condition & Efficiency & Purity & Mean $\chi^2$ \\ \hline  
$W$ mass known exactly & & & \\ 
$t$ pole mass assumed & \raisebox{1.5ex}[-1.5ex]{0.893} & \raisebox{1.5ex}[-1.5ex]{$5.21\cdot 10^{-5}$} & \raisebox{1.5ex}[-1.5ex]{417.3} \\ \hline 
$t,W$ pole mass assumed & 0.839 & 0 & 911.0 \\ \hline 
$t,W$ pole mass assumed & & &  \\ 
both $b \bar{b}$ permutations & \raisebox{1.5ex}[-1.5ex]{0.890} & \raisebox{1.5ex}[-1.5ex]{0} & \raisebox{1.5ex}[-1.5ex]{1166} \\ \hline 
reconstructed $b$-jets & & & \\
(parton matched) & \raisebox{1.5ex}[-1.5ex]{0.711} & \raisebox{1.5ex}[-1.5ex]{0} & \raisebox{1.5ex}[-1.5ex]{3916} \\ \hline
wrong $b$-jet permutation & & & \\
(parton matched) & \raisebox{1.5ex}[-1.5ex]{0.426} & \raisebox{1.5ex}[-1.5ex]{0} & \raisebox{1.5ex}[-1.5ex]{5366} \\ \hline
both $b$-jet permutations & & & \\
(parton matched) & \raisebox{1.5ex}[-1.5ex]{0.822} & \raisebox{1.5ex}[-1.5ex]{0} & \raisebox{1.5ex}[-1.5ex]{4049} \\ \hline
both $b$-jet permutations & & & \\
(parton matched, & 0.761 & 0 & 4491 \\ 
 jets + leptons smeared) & & & \\ \hline
both $b$-jet permutations & & & \\
(parton matched, & & & \\ 
 jets + leptons smeared), & 0.994 & 0 & 2556 \\ 
reconstructed objects & & & \\
100 times resolution smeared & & & \\ \hline

\end{tabular}
\caption{\label{solution_characteristics} Solutions fulfilling $\chi^2<10^{-3}$
are defined as matched and else unmatched. The purity is determined according to this definition. The Mean $\chi^2$ is obtained taking into account matched and unmatched events.
}
\end{table}
The solutions coincide in 99.7\% of cases within real precision
to the generated neutrino momenta.
Fig. \ref{DataSolChi2} shows impressively how accurate and reliably 
the method is working. 
The plots in fig. \ref{DataUnmatchEvtLeastSolChi2}
show the $\chi^2$ distribution on a linear scale.
Since in practice the off-shell masses of the top quark and $W$ boson
resonances are not known the method has been applied in the following ways:
The distribution in the first plot assumes $W$ boson off-shell but
top quark pole masses. It peaks at zero and its tail vanishes rapidly.
The solution efficiency for this scenario amounts to 89\%.
The second plot assumes the 
pole mass for the top quarks and the $W$ bosons. The number
and mean of unmatched solutions increases dramatically and 
the efficiency drops to 84\%.
Further an infrared-safe cone algorithm \cite{H11999} with cone size $R=0.5$
in the space spanned by pseudorapidity and azimuthal angle 
has been applied to the hadronic final state particles to investigate the 
effect of reconstructed objects on the solutions.
Requiring exactly two jets and two leptons 
and accepting the jets as $b$-tagged if they coincide within $\Delta R<0.5$
with the $b$ quarks and antiquarks yields an significant degradation of the $\chi^2$ 
distribution (fig. \ref{DataUnmatchEvtLeastSolChi2} c)).
The efficiency drops to 43\% assuming the right jet quark combination. 
Admitting both permutations yields an efficiency of 82\%.
The last plot has been obtained from the previous one in additionally
smearing the leptons and jets with the energy resolution 
of the D$\emptyset$ detector \cite{D0note4677}.
The $\chi^2$ of the minimal solution suffers in average another ten percent
degradation and the solution efficiency drops by the same amount.  
In practice a given event passes the solving procedure repeatedly to improve the
solution efficiency. Each iteration the energy of the reconstructed objects is
randomly drawn from a probability distribution describing the detector resolution
and centred around the measured values. 
In the case of hundred such iterations the efficiency can be kept above 99.4\% 
while in average the $\chi^2$ of the best solution decreases considerably
as expected in comparison to solving the momenta of the reconstructed objects just once.

In table \ref{solution_characteristics} the efficiencies and minimal solution $\chi^2$'s
are summarised. In addition the purity is given. It is practically zero, which means
that no solutions do match with real precision or even merely with a $\chi^2$ better 
than $10^{-3}$ once the off-shell masses of the top quarks and $W$ bosons 
are not assumed to be known exactly.

General numerical methods can compare and gauge their performance in terms of solution
efficiency and purity with the algebraic approach described here.


The time consumption of the method amounts to about 20\% of the time needed
for the generation of the events which means if $5\cdot 10^6$ events can
be generated in five hours an additional hour is needed to solve them.
The strength of the method is the application of Sturm's theorem, such that
in the case of no solutions the time consuming seeking and polishing of 
solutions can be saved. The bottleneck of the method is the time consuming
evaluation of the resultant.

\section{Conclusions}

\noindent
An algebraic approach to solve the $t\bar{t}$ dilepton kinematics
has been presented. The system of equations can be reduced to a 
univariate polynomial by means of resultants.
The number of real roots can be determined by means of Sturm's theorem.
Once the single roots have been isolated they can be polished 
by binary bracketing while seeking for the sign change.
In this way a solution is found in 99.9\% of cases.
The solutions coincide with real precision to the generated neutrino 
energies and momenta in 99.7\% of cases assuming that the reconstructible
parton momenta inserted in the solving procedure are known exactly.
Little deviations drop the solution efficiency considerably, 
at the order of tens of percent. In this case the solved neutrino momenta 
differ already in average by the order of tens of GeV from the generated
parton momenta. 
The solution efficiency can be re-established above the 99\% level
in solving a given event several times, varying the energy of the reconstructed 
objects each iteration randomly according to the energy resolution of a detector. 
General numerical methods can compare their performance in terms of efficiency 
and purity to the algebraic approach whose implementation has been described here.

\begin{acknowledgments}

\noindent
Many thanks to Paul Russo for useful discussions how to implement
the code in a very efficient manner. I'm also grateful to many colleagues of
{\it Universit\'es de Paris VI, VII} and the D$\emptyset$ collaboration
for helpful suggestions and proof-reading of the manuscript. 
This work has been supported by the 
{\it Commissariat \`a l'Energie Atomique} and
CNRS/{\it{Institut National de Physique Nucl\'eaire et de Physique des
Particules}}, France.
Many thanks to Bruno Wittmer and Georgios Anagnostou for useful discussions.

\end{acknowledgments}

\appendix

\section*{Coefficients}

\noindent
Before defining the coefficients of equations (\ref{pzcoefficients}) and 
(\ref{pzbarcoefficients}) it is useful to introduce the following
invariant masses
\begin{eqnarray} \nonumber
 m_{\ell^+} & = & \sqrt{E_{\ell^+}^2-p_{\ell^+_x}^2-p_{\ell^+_y}^2-p_{\ell^+_z}^2} \\ \nonumber
 m_{\ell^-} & = & \sqrt{E_{\ell^-}^2-p_{\ell^-_x}^2-p_{\ell^-_y}^2-p_{\ell^-_z}^2} \\ \nonumber
 m_b & = & \sqrt{E_b^2-p_{b_x}^2-p_{b_y}^2-p_{b_z}^2} \\ \nonumber
 m_{\bar{b}} & = & \sqrt{E_{\bar{b}}^2-p_{\bar{b}_x}^2-p_{\bar{b}_y}^2-p_{\bar{b}_z}^2} \\ \nonumber
 m_{b\ell^+} & = & \left\{(E_b+E_{\ell^+})^2-(p_{b_x}+p_{\ell^+_x})^2 \right. \\ \nonumber
& & \left. -(p_{b_y}+p_{\ell^+_y})^2-(p_{b_z}+p_{\ell^+_z})^2 \right\}^{\frac{1}{2}} \\ \nonumber
 m_{\bar{b}\ell^-} & = & \left\{(E_{\bar{b}}+E_{\ell^-})^2-(p_{\bar{b}_x}+p_{\ell^-_x})^2 \right. \\ \nonumber
 & & \left. -(p_{\bar{b}_y}+p_{\ell^-_y})^2-(p_{\bar{b}_z}+p_{\ell^-_z})^2 \right\}^{\frac{1}{2}} \\ \nonumber
\end{eqnarray}

The coefficients are then given by 
\begin{eqnarray} \nonumber
 a_{11} & = & \frac{1}{2}\frac{(m_{W^+}^2-m_{\ell^+}^2-m_{\nu}^2)p_{\ell^+_z}}{E_{\ell^+}^2-p_{\ell^+_z}^2} \\ \nonumber
 a_{12} & = & \frac{p_{\ell^+_x}p_{\ell^+_z}}{E_{\ell^+}^2-p_{\ell^+_z}^2} \\ \nonumber
 a_{13} & = & \frac{p_{\ell^+_y}p_{\ell^+_z}}{E_{\ell^+}^2-p_{\ell^+_z}^2} \\ \nonumber
 a_{21} & = & \frac{1}{4(E_{\ell^+}^2-p_{\ell^+_z}^2)}\cdot\left( m_{W^+}^4+m_{\ell^+}^4+m_{\nu}^4 \right. \\ \nonumber
 & & \left. -2m_{W^+}^2m_{\ell^+}^2-2m_{W^+}^2m_{\nu}^2+2m_{\ell^+}^2m_{\nu}^2 \right) \\ \nonumber
 a_{22} & = & \frac{(m_{W^+}^2-m_{\ell^+}^2-m_{\nu}^2)p_{\ell^+_x}}{E_{\ell^+}^2-p_{\ell^+_z}^2} \\ \nonumber
 a_{23} & = & \frac{(m_{W^+}^2-m_{\ell^+}^2-m_{\nu}^2)p_{\ell^+_y}}{E_{\ell^+}^2-p_{\ell^+_z}^2} \\ \nonumber
 a_{24} & = & -\frac{E_{\ell^+}^2-p_{\ell^+_x}^2}{E_{\ell^+}^2-p_{\ell^+_z}^2} \\ \nonumber
 a_{25} & = & \frac{2p_{\ell^+_x}p_{\ell^+_y}}{E_{\ell^+}^2-p_{\ell^+_z}^2} \\ \nonumber
 a_{26} & = & -\frac{E_{\ell^+}^2-p_{\ell^+_y}^2}{E_{\ell^+}^2-p_{\ell^+_z}^2} \; . \\ \nonumber
\end{eqnarray}
To obtain the coefficients $c_{mn}$ for the other parton branch
one has to substitute $W^+$, $\ell^+$ and $\nu$ by $W^-$, $\ell^-$ and $\bar{\nu}$ respectively. 
Similar holds
\begin{eqnarray} \nonumber
 b_{11} & = & \frac{1}{2}\frac{(m_t^2-m_{b\ell^+}^2-m_{\nu}^2)(p_{b_z}+p_{\ell^+_z})}{(E_b+E_{\ell^+})^2-(p_{b_z}+p_{\ell^+_z})^2} \\ \nonumber
 b_{12} & = & \frac{(p_{b_x}+p_{\ell^+_x})(p_{b_z}+p_{\ell^+_z})}{(E_b+E_{\ell^+})^2-(p_{b_z}+p_{\ell^+_z})^2} \\ \nonumber
 b_{13} & = & \frac{(p_{b_y}+p_{\ell^+_y})(p_{b_z}+p_{\ell^+_z})}{(E_b+E_{\ell^+})^2-(p_{b_z}+p_{\ell^+_z})^2} \\ \nonumber
 b_{21} & = & \frac{1}{4((E_b+E_{\ell^+})^2-(p_{b_z}\!\!+p_{\ell^+_z})^2)} \cdot \left( m_t^4+m_{b\ell^+}^4\!+m_{\nu}^4 \right. \\ \nonumber
        &  & \left. -2m_{t}^2m_{b\ell^+}^2-2m_{t}^2m_{\nu}^2+2m_{b\ell^+}^2m_{\nu}^2 \right) \\ \nonumber
 b_{22} & = & \frac{(m_t^2-m_{b\ell^+}^2-m_{\nu}^2)(p_{b_x}+p_{\ell^+_x})}{(E_b+E_{\ell^+})^2-(p_{b_z}+p_{\ell^+_z})^2} \\ \nonumber
 b_{23} & = & \frac{(m_t^2-m_{b\ell^+}^2-m_{\nu}^2)(p_{b_y}+p_{\ell^+_y})}{(E_b+E_{\ell^+})^2-(p_{b_z}+p_{\ell^+_z})^2} \\ \nonumber
 b_{24} & = & -\frac{(E_b+E_{\ell^+})^2-(p_{b_x}+p_{\ell^+_x})^2}{(E_b+E_{\ell^+})^2-(p_{b_z}+p_{\ell^+_z})^2} \\ \nonumber
 b_{25} & = & \frac{2(p_{b_x}+p_{\ell^+_x})(p_{b_y}+p_{\ell^+_y})}{(E_b+E_{\ell^+})^2-(p_{b_z}+p_{\ell^+_z})^2} \\ \nonumber
 b_{26} & = & -\frac{(E_b+E_{\ell^+})^2-(p_{b_y}+p_{\ell^+_y})^2}{(E_b+E_{\ell^+})^2-(p_{b_z}+p_{\ell^+_z})^2} \; . \\ \nonumber
\end{eqnarray}
Again the coefficients $d_{mn}$ of the other parton branch can be obtained 
in substituting $t$, $b$, $\ell^+$ and $\nu$ by $\bar{t}$, $\bar{b}$, $\ell^-$ and $\bar{\nu}$
respectively. The denominators are always of the type 
\begin{eqnarray} \nonumber
E^2-p_z^2 = m^2 + p_{\perp}^2 \geq m^2 \; .
\end{eqnarray}
Thus it is ensured that they never vanish. Running over 1 million Monte-Carlo 
events does not lead to a division by zero.
In addition, detected objects in collider experiments have always a considerable 
amount of transverse momentum which pushes the kinematics of the equations 
further away from such singularities. Therefore the theoretically possible 
multiplication of all equations with the least common multiple of all 
denominators does not need to be applied.

\bibliography{apssamp}

\pagebreak

\end{document}